\pdfoutput=1
\PassOptionsToPackage{table}{xcolor} 
\documentclass[11pt]{article}

\usepackage[]{ACL2023}

\usepackage{times}
\usepackage{latexsym}
\usepackage[T1]{fontenc}
\usepackage[utf8]{inputenc}
\usepackage{inconsolata}

\usepackage{lipsum}
\usepackage{graphicx}
\usepackage{microtype}
\usepackage{hyperref}
\usepackage{url}
\usepackage{booktabs, nicematrix}
\usepackage[table]{xcolor} 
\usepackage{multirow,makecell}
\usepackage{amsmath, amsmath}
\usepackage{xspace}
\usepackage{wrapfig}
\usepackage{pifont}
\usepackage{paralist}
\usepackage{enumitem}

\newcommand{\cmark}{\ding{51}}%
\newcommand{\xmark}{\ding{55}}%

\usepackage{arydshln}
\usepackage{listings}
\usepackage{adjustbox}
\usepackage[most]{tcolorbox}

\definecolor{dark-gray}{gray}{0.85}
\definecolor{light-gray}{gray}{0.95}
\definecolor{mygreen}{rgb}{0,0.4,0}
\definecolor{mygray}{rgb}{0.5,0.5,0.5}
\definecolor{mymauve}{rgb}{0.58,0,0.82}
\definecolor{myred}{rgb}{0.82, 0.1, 0.26}

\lstdefinestyle{CustomPy}{
    escapeinside={(*@}{@*)},
    belowcaptionskip=1\baselineskip,
    xleftmargin=1pt,
    xrightmargin=1pt,
    language=Python,
    numbersep=5pt,
    tabsize=4,
    showstringspaces=false,
    basicstyle=\small\ttfamily, 
    keywordstyle=\bf\color{mygreen},
    commentstyle=\color{purple},
    stringstyle=\color{red},
    identifierstyle=\color{black},
    numberstyle=\tiny\color{mygray},
    emph={int,char,double,float,unsigned,void,bool,boolean},
    emphstyle={\bf\color{myred}},
    emph=[2]{and, in,},
    emphstyle=[2]{\bf\color{violet}},
    emph=[3]{even_odd_count, single_line_infill_postprocess, multi_line_infill_postprocess, random_span_infill_postprocess, remove_overlap_prefix_middle, remove_overlap_middle_suffix},
    emphstyle=[3]{\bf\color{blue}},
    numbers=left,
    stepnumber=1,
    breaklines=true,
    backgroundcolor=\color{white},
    literate={\ \ }{{\ }}1,
}

\makeatletter
\let\old@lstKV@SwitchCases\lstKV@SwitchCases
\def\lstKV@SwitchCases#1#2#3{}
\makeatother
\usepackage{lstlinebgrd}
\makeatletter
\let\lstKV@SwitchCases\old@lstKV@SwitchCases

\lst@Key{numbers}{none}{%
    \def\lst@PlaceNumber{\lst@linebgrd}%
    \lstKV@SwitchCases{#1}%
    {none:\\%
     left:\def\lst@PlaceNumber{\llap{\normalfont
                \lst@numberstyle{\thelstnumber}\kern\lst@numbersep}\lst@linebgrd}\\%
     right:\def\lst@PlaceNumber{\rlap{\normalfont
                \kern\linewidth \kern\lst@numbersep
                \lst@numberstyle{\thelstnumber}}\lst@linebgrd}%
    }{\PackageError{Listings}{Numbers #1 unknown}\@ehc}}
\makeatother

\newcount\myloopcounter

\newcommand{\repeatit}[2][10]{%
  \myloopcounter0
  \loop\ifnum\myloopcounter < #1 
  #2%
  \advance\myloopcounter by 1 %
  \repeat 
}

\title{From Output to Evaluation: Does Raw Instruction-Tuned Code LLMs Output Suffice for Fill-in-the-Middle Code Generation?}

\author{
Wasi Uddin Ahmad, Somshubra Majumdar, Boris Ginsburg \\ [1pt]
NVIDIA\\ [1pt]
Santa Clara, CA 95051, USA \\ [1pt]
\texttt{\{wasiuddina, smajumdar\}@nvidia.com}
}

\begin{document}
\maketitle
\begin{abstract}
Post-processing is crucial for the automatic evaluation of LLMs in fill-in-the-middle (FIM) code generation due to the frequent presence of extraneous code in raw outputs. This extraneous generation suggests a lack of awareness regarding output boundaries, requiring truncation for effective evaluation. The determination of an optimal truncation strategy, however, often proves intricate, particularly when the scope includes several programming languages. This study investigates the necessity of post-processing instruction-tuned LLM outputs. Our findings reveal that supervised fine-tuning significantly enhances FIM code generation, enabling LLMs to generate code that seamlessly integrates with the surrounding context.
Evaluating our fine-tuned \texttt{Qwen2.5-Coder} (base and instruct) models on HumanEval Infilling and SAFIM benchmarks demonstrates improved performances without post-processing, especially when the \emph{middle} consists of complete lines. However, post-processing of the LLM outputs remains necessary when the \emph{middle} is a random span of code.
\end{abstract}

\section{Introduction}

The iterative process of coding, involving frequent edits and insertions \cite{bavarian2022efficient, fried2023incoder}, establishes Fill-in-the-Middle (FIM) code generation a prevalent task in code completion. Models tackling this must generate the missing code segment conditioned on both the preceding (left) and succeeding (right) context. A key challenge in FIM lies in seamlessly integrating the generated \emph{middle} with the subsequent code while maintaining both structure and meaning – a non-trivial learning objective for models. Consequently, raw model outputs often undergo rule-based post-processing to remove extraneous content. As shown in \autoref{tab:postprocess}, two widely used FIM code generation evaluation benchmarks employ specific truncation rules that may not generalize to real-world FIM scenarios with arbitrary left and right contexts. Furthermore, such truncation strategies often fail to account for alternative, yet valid, ways of generating the missing code. For instance, as illustrated in \autoref{fig:example1}, a single-line infilling task might expect one line as a solution, but an LLM could generate five lines that perfectly match the surrounding context. In this case, truncating the generated middle to a single line would incorrectly mark it as a failure.
Given the advancements in code LLMs, a crucial question emerges: do modern code LLMs naturally know when to stop generating given any arbitrary left and right context, thereby eliminating the need for post-processing techniques like truncation?

\begin{table*}[t]
\centering
\begin{NiceTabular}{l| l| X[l]}
\toprule
\textbf{Dataset} & \textbf{FIM Type} & \textbf{Truncation Rule for Output} \\
\midrule
\multirowcell{3}[0pt][l]{HumanEval Infilling} & Single-line & Truncate output to one line. \\
& Multi-line & Truncate output to match ground truth line count. \\
& Random-span & Truncate output if overlaps with prefix and suffix. \\
\midrule
\multirowcell{3}[0pt][l]{SAFIM} & Algorithm-block & Truncate output to one line. \\
& Control-flow & Truncate output to match ground truth program structure. \\
& API function call & Truncate output after first closing parenthesis. \\
\bottomrule
\end{NiceTabular}
\caption{Truncation strategy used in two popular FIM benchmarks.
}
\label{tab:postprocess}
\end{table*}

\begin{figure*}[!htb]
\centering
\begin{adjustbox}{valign=t,minipage=0.4\textwidth}
\begin{center}
    \underline{Canonical solution}
\end{center} 
\begin{tabular}{l}
\lstset{escapechar=@,style=CustomPy}
\begin{lstlisting}[ 
    linebackgroundcolor={%
    \ifnum\value{lstnumber}=5
        \color{red!10}
    \fi
    }
]
def even_odd_count(num):
    """[docstring truncated]"""
    even_count = 0
    odd_count = 0
    for i in str(abs(num)):       
        if int(i)%2==0:
            even_count +=1
        else:
            odd_count +=1
    return (even_count, odd_count)
\end{lstlisting}
\end{tabular}
\end{adjustbox}
\hspace{30pt}
\begin{adjustbox}{valign=t,minipage=0.4\textwidth}
\begin{center}
    \underline{FIM by our finetuned model}
\end{center} 
\begin{tabular}{l}
\lstset{escapechar=@,style=CustomPy}
\begin{lstlisting}[ 
    linebackgroundcolor={%
    \ifnum\value{lstnumber}>4
        \ifnum\value{lstnumber}<10
            \color{green!10}
        \fi
    \fi
    }
]
def even_odd_count(num):
    """[docstring truncated]"""
    even_count = 0
    odd_count = 0
    if num < 0:
        num = str(num)[1:]
    else:
        num = str(num)
    for i in num:      
        if int(i)%2==0:
            even_count +=1
        else:
            odd_count +=1
    return (even_count, odd_count)
\end{lstlisting}
\end{tabular}
\end{adjustbox}
\caption{
{\bf Left}: An example of a single-line infilling task (highlighted in red) from the HumanEval Infilling benchmark. {\bf Right}: A fill-in-the-middle generation produced by our fine-tuned Qwen2.5-Coder-7B-Instruct model.
}
\vspace{-2mm}
\label{fig:example1}
\end{figure*}


The existing body of work \cite{bavarian2022efficient, fried2023incoder, nguyen2023meet, zheng2024selfinfilling} predominantly examines \emph{base} LLMs, trained on massive amounts of data to understand language patterns and generate consistent output. These models acquire Fill-in-the-Middle (FIM) capabilities by learning from reordered prefix-middle-suffix sequences, created via random splits of the training data. The purpose of this reordering is to allow the LLM to auto-regressively predict the middle segment, conditioned on both the left and right contexts as past information. In contrast to base LLMs, we posit that \emph{instruction-tuned} LLMs are better equipped for FIM generation due to their customized nature and their inherent capacity to adhere to instructions. 
Our primary motivation for focusing on instruction-tuned LLMs stems from the objective to avoid the expensive pre-training (or their continuation) required by models like those in \citep{bavarian2022efficient}, which demonstrated that fine-tuning with FIM does not achieve the same performance as pre-training with FIM.

This study investigates the necessity of post-processing instruction-tuned LLM outputs for FIM code generation. Our empirical analysis reveal that the raw outputs of off-the-shelf instruction-tuned LLMs often require editing. Consequently, we fine-tuned both base and instruct versions of \texttt{Qwen2.5-Coder}. Our findings demonstrate that these fine-tuned models can produce outputs that do not require any post-processing when the middle code segments consist of whole lines. In fact, applying any preset, heuristic-based post-processing in such cases actually leads to incorrect middle outputs. However, when middle segments comprise partial lines, it becomes necessary to truncate overlapping code segments. Based on our findings, we offer straightforward post-processing recommendations for LLM-generated middle code segments.

In summary, we contribute the followings:
\begin{compactenum}
    \item We show that off-the-shelf instruction-tuned LLMs require post-processing for effective FIM code generation and exhibit suboptimal performance due to a lack of task-specific finetuning or optimization.
    \item We demonstrate that lightweight fine-tuning significantly boosts LLM performance for FIM generation. Interestingly, when the \emph{middle} code consists of complete lines, the raw outputs from these fine-tuned models achieve better automatic evaluation scores than post-processed outputs, meaning no further editing is needed. However, if the \emph{middle} includes partial lines, post-processing is still required.
\end{compactenum}

\section{Instruction-tuning of LLMs for Fill-in-the-Middle Code Generation}

We investigate the FIM code generation accuracy of state-of-the-art instruction-tuned code LLMs by prompting them with instructions, as illustrated in \autoref{fig:utg_prompt}. This prompting method is consistent with their standard usage for code generation. Our findings in \autoref{sec:results} reveal that instruction-tuned LLMs perform suboptimally, even after their outputs undergo dataset-specific post-processing. Building on this observation, we further investigate if lightweight supervised fine-tuning can empower code LLMs for improved FIM generation.

To achieve this, we created a training dataset of instruction-response pairs using an LLM. First, we collected a set of Python functions from GitHub, following the data collection pipeline detailed in \citet{wei2024selfcodealign}. This involved a rigorous filtering process: type checking with Pyright, removal of benchmark items, elimination of poorly documented functions, and deduplication.
Using these collected functions, we employed a straightforward approach to generate instruction-response pairs. Specifically, we prompted Mixtral-8x22B \cite{jiang2024mixtralexperts} with the template shown in \autoref{fig:datagen}, asking it to split each function into prefix, middle, and suffix according to one of five strategies outlined in the prompt.
After generating the prefix, middle, and suffix, we verified that their concatenation reconstructs the original function. At the end, we collected $\approx$1M instruction-response pairs that we used to finetune code LLMs.

\section{Experiments}
\subsection{Setup}

\begin{table*}[t]
\centering
\resizebox{1.0\linewidth}{!} {%
\def\arraystretch{1.0}%
{\setlength{\tabcolsep}{8pt}
\begin{NiceTabular}{l |c|c c c c|c c c c}
\toprule
\multirow{2}{*}{\bf Model} & {\bf Post-} & \multicolumn{4}{c|}{\bf HumanEval Infilling} & \multicolumn{4}{c}{\bf SAFIM (Python)} \\
& {\bf Proc.} & SL & ML & RS & Avg. & Algo. & Control & API & Avg. \\

\midrule

\multirow{2}{*}{Qwen2.5-Coder-7B-Instruct}  & \xmark & 47.1 & 22.4 & 0.9 & 23.5 & 3.5 & 0	& 0 & 1.2 \\
& \cmark & 53.3 & 24.3 & 12.4 & 30.0 & 3.98 & 14.4 & 22.1 & 13.5 \\
\hdashline

\cellcolor{gray!10} & 
\cellcolor{gray!10} \xmark & 
\cellcolor{gray!10} 89.3 & 
\cellcolor{gray!10} 68.2 & 
\cellcolor{gray!10} 31.9 & 
\cellcolor{gray!10} 63.1 & 
\cellcolor{gray!10} 28.5 & 
\cellcolor{gray!10} 29.1 & 
\cellcolor{gray!10} 69.6 & 
\cellcolor{gray!10} 42.4 \\

\multirow{-2}{*}{\cellcolor{gray!10}Qwen2.5-Coder-7B} & 
\cellcolor{gray!10} \cmark & 
\cellcolor{gray!10} 85.7 & 
\cellcolor{gray!10} 61.3 & 
\cellcolor{gray!10} {\bf 43.7} & 
\cellcolor{gray!10} 63.6 & 
\cellcolor{gray!10} 28.3 & 
\cellcolor{gray!10} {\bf 36.7} & 
\cellcolor{gray!10} 69.1 & 
\cellcolor{gray!10} 44.7 \\ 
\hdashline

\cellcolor{gray!25} & 
\cellcolor{gray!25} \xmark & 
\cellcolor{gray!25} {\bf 91.6} & 
\cellcolor{gray!25} {\bf 67.4} & 
\cellcolor{gray!25} 34.2 & 
\cellcolor{gray!25} {\bf 64.4} & 
\cellcolor{gray!25} 30.7 & 
\cellcolor{gray!25} 30.1 & 
\cellcolor{gray!25} {\bf 72.4} & 
\cellcolor{gray!25} 44.4 \\

\multirow{-2}{*}{\cellcolor{gray!25}Qwen2.5-Coder-7B-Instruct} & 
\cellcolor{gray!25} \cmark & 
\cellcolor{gray!25} 88.7 & 
\cellcolor{gray!25} 61.0 & 
\cellcolor{gray!25} 43.0 & 
\cellcolor{gray!25} 64.2 & 
\cellcolor{gray!25} {\bf 30.8} & 
\cellcolor{gray!25} 36.0 & 
\cellcolor{gray!25} 69.6 & 
\cellcolor{gray!25} {\bf 45.5} \\

\midrule

\multirow{2}{*}{Qwen2.5-Coder-14B-Instruct} & \xmark & 43.9 & 27.8 & 3.4 & 25.0 & 7.2 & 0 & 2.2 & 3.1 \\
& \cmark & 46.7 & 27.4 & 12.2 & 28.8 & 12.7 & 15.1 & 43.7 & 23.8 \\
\hdashline

\cellcolor{gray!10} & 
\cellcolor{gray!10} \xmark & 
\cellcolor{gray!10} 87.0 & 
\cellcolor{gray!10} 72.7 & 
\cellcolor{gray!10} 36.4 & 
\cellcolor{gray!10} 65.4 & 
\cellcolor{gray!10} 23.7 & 
\cellcolor{gray!10} 29.6 & 
\cellcolor{gray!10} 74.0 & 
\cellcolor{gray!10} 42.4 \\

\multirow{-2}{*}{\cellcolor{gray!10}Qwen2.5-Coder-14B} &
\cellcolor{gray!10} \cmark & 
\cellcolor{gray!10} 84.9 & 
\cellcolor{gray!10} 63.9 & 
\cellcolor{gray!10} 46.3 & 
\cellcolor{gray!10} 65.0 & 
\cellcolor{gray!10} 23.7 & 
\cellcolor{gray!10} 35.5 & 
\cellcolor{gray!10} 72.9 & 
\cellcolor{gray!10} 44.0 \\ 
\hdashline

\cellcolor{gray!25} & 
\cellcolor{gray!25} \xmark & 
\cellcolor{gray!25} {\bf 91.7} & 
\cellcolor{gray!25} {\bf 73.4} & 
\cellcolor{gray!25} 37.6 & 
\cellcolor{gray!25} {\bf 67.6} & 
\cellcolor{gray!25} 29.4 & 
\cellcolor{gray!25} 33.5 & 
\cellcolor{gray!25} {\bf 76.8} & 
\cellcolor{gray!25} 46.6 \\

\multirow{-2}{*}{\cellcolor{gray!25}Qwen2.5-Coder-14B-Instruct} & 
\cellcolor{gray!25} \cmark & 
\cellcolor{gray!25} 88.8 & 
\cellcolor{gray!25} 64.3 & 
\cellcolor{gray!25} {\bf 46.8} & 
\cellcolor{gray!25} 66.6 & 
\cellcolor{gray!25} {\bf 29.8} & 
\cellcolor{gray!25} {\bf 38.7} & 
\cellcolor{gray!25} 75.1 & 
\cellcolor{gray!25} {\bf 47.9} \\

\midrule

\multirow{2}{*}{Qwen2.5-Coder-32B-Instruct} & \xmark & 74.7 & 47.7 & 5.9 & 42.8 & 19 & 1.7 & 10 & 10.2 \\
& \cmark & 77.0 & 56.9 & 22.7 & 52.2 & 19.5 & 25.3 & 45.9 & 30.2 \\
\hdashline

\cellcolor{gray!10} & 
\cellcolor{gray!10} \xmark & 
\cellcolor{gray!10} 93.9 & 
\cellcolor{gray!10} 75.3 & 
\cellcolor{gray!10} 36.6 & 
\cellcolor{gray!10} 68.6 & 
\cellcolor{gray!10} 31.4 & 
\cellcolor{gray!10} 36.1 & 
\cellcolor{gray!10} 74.6 & 
\cellcolor{gray!10} 47.4 \\

\multirow{-2}{*}{\cellcolor{gray!10}Qwen2.5-Coder-32B} & 
\cellcolor{gray!10} \cmark & 
\cellcolor{gray!10} 89.7 & 
\cellcolor{gray!10} 66.8 & 
\cellcolor{gray!10} 47.9 & 
\cellcolor{gray!10} 68.1 & 
\cellcolor{gray!10} {\bf 31.8} & 
\cellcolor{gray!10} {\bf 41.7} & 
\cellcolor{gray!10} 75.1 & 
\cellcolor{gray!10} {\bf 49.5} \\ 
\hdashline

\cellcolor{gray!25} &
\cellcolor{gray!25} \xmark & 
\cellcolor{gray!25} {\bf 94.8} & 
\cellcolor{gray!25} {\bf 76.5} & 
\cellcolor{gray!25} 37.6	& 
\cellcolor{gray!25} {\bf 69.6} & 
\cellcolor{gray!25} 31.6 & 
\cellcolor{gray!25} 36.7 & 
\cellcolor{gray!25} {\bf 76.2} & 
\cellcolor{gray!25} 48.2 \\

\multirow{-2}{*}{\cellcolor{gray!25}Qwen2.5-Coder-32B-Instruct} & 
\cellcolor{gray!25} \cmark &	
\cellcolor{gray!25} 91.4 & 
\cellcolor{gray!25} 68.7 & 
\cellcolor{gray!25} {\bf 48.0} & 
\cellcolor{gray!25} 69.4 & 
\cellcolor{gray!25} 31.6 & 
\cellcolor{gray!25} {\bf 41.7} & 
\cellcolor{gray!25} 74.6 & 
\cellcolor{gray!25} 49.3 \\

\bottomrule
\end{NiceTabular}
}
}
\caption{Performance comparison of Qwen2.5-Coder-Instruct models across three different sizes. SL, ML, and RS indicate ``single-line'', ``multi-line'', and ``random-span'' infilling tasks, respectively. Highlighted rows show our finetuned models' performances. Bold indicates the highest performances for each model groups.}
\label{tab:main_results}
\vspace{-2mm}
\end{table*}

\paragraph{Training \& Inference Setup}
We fine-tuned the 7B, 14B, and 32B parameter base and instruct versions of \texttt{Qwen2.5-Coder}. 
The finetuning spanned 5000 steps on NVIDIA H100-80GB GPUs, leveraging the AdamW optimizer \citep{kingma2014adam} with a batch size of 256 and a maximum sequence length of 4096 tokens.
We initialized the learning rate at 5e-6 and applied a CosineAnnealing scheduler with a 10\% warmup. We utilized tensor parallelism and BF16 precision to accelerate the training process. For evaluation, we utilized the final training checkpoint, and during inference, we employed greedy decoding.

\paragraph{Evaluation Benchmarks and Metrics}
We evaluated models using two FIM code generation benchmarks: HumanEval Infilling \cite{bavarian2022efficient} and SAFIM \cite{gong2024evaluation}. The HumanEval Infilling benchmark features three distinct tasks: Single-line, Multi-line, and Random span infilling. We provide the post-processing functions for these tasks in \autoref{fig:code_example}.
In contrast, SAFIM is a syntax-aware FIM benchmark, consisting of tasks focused on algorithm block, control flow, and API function call completion. For both benchmarks, we present results based on the standard pass@1 metric.\footnote{For SAFIM evaluation, we used the authors released code, available at \url{https://github.com/gonglinyuan/safim}.}

\subsection{Research Questions}

We aim to address the following questions.
\begin{compactenum}
    \item What is the out-of-the-box effectiveness of instruction-tuned code LLMs for fill-in-the-middle (FIM) code generation?
    \item Can supervised fine-tuning significantly improve the FIM generation accuracy of code LLMs? How does finetuning impact \emph{base} and \emph{instruct} version of LLM? Furthermore, what are the effects of such fine-tuning on \emph{base} vs. \emph{instruct}-tuned LLMs?
    \item Are the raw outputs of fine-tuned LLMs sufficiently effective for automatic evaluation?
\end{compactenum}

\subsection{Results}
\label{sec:results}
The results are presented in \autoref{tab:main_results}. We consistently observed a few performance trends.

\paragraph{Instruction-tuned LLMs are not ready out-of-the-box} 
The \texttt{Qwen2.5-Coder-Instruct} models consistently perform poorly on both benchmarks, particularly on the SAFIM and random-span HumanEval infilling tasks. Their low accuracies clearly indicate that these models cannot be effectively used off-the-shelf in FIM generation.

\paragraph{Supervised finetuning (SFT) is a major leap for FIM instruction-tuning} 
The overall results clearly indicate a significant performance boost with SFT of \texttt{Qwen2.5-Coder-Instruct} models. The average performance of the 7B and 14B models doubled, while the 32B models saw an impressive 40-50\% improvement compared to their of-the-shelf counterparts.

\paragraph{Sample efficiency of base vs. instruct LLMs} 
The average pass@1 accuracies across both benchmarks suggest that tuning instruction-following LLMs yields slightly better performance.

\paragraph{Raw outputs of finetuned LLMs are effective} 
From \autoref{tab:main_results}, we observe that post-processing consistently lowers accuracies for single-line and multi-line infilling tasks in HumanEval benchmark as shown in \autoref{fig:example1}. However, for random-span infilling, raw LLM outputs do require editing, which is evident from the resulting improved performance after post-processing. We see a similar pattern for the SAFIM benchmark.

Based on our observations and experiment results, we recommend \textit{post-processing to remove overlapping code segments found between the prefix and the generated middle, and similarly between the middle and the suffix}. This is our standard approach for all infilling tasks in this work.

\subsection{Other Findings}
Our experiments showed that generating multiple FIM samples from a single Python function (resulting in 5M instruct-response pairs) didn't significantly improve supervised fine-tuning. Thus, we suggest future work prioritize diversity in Python functions over generating many samples from one. 

Additionally, fine-tuning models for more than roughly one epoch degraded performance on downstream FIM tasks. Therefore, we recommend using a larger collection of training samples, but with only a single training iteration over them.

\section{Related Work}

\citet{bavarian2022efficient} presented a foundational approach to training large language models (LLMs) for FIM code generation, marking a significant first step in this area. 
Their core innovation involved segmenting unlabeled code into three distinct parts and rearranging those segments to create training sequences. 
This pioneering strategy proved highly influential, shaping nearly all subsequent research in FIM code generation \cite{fried2023incoder, zheng2024selfinfilling, wu2024repoformer, sagtani2025improving}.

In contrast to this dominant paradigm, \citet{nguyen2023meet} introduced an alternative method. They trained two separate language models, each generating code in an opposing direction: one from left-to-right and the other from right-to-left. The FIM task was then solved by having these independently generated segments converge and ``meet'' in the middle. More recently, \citet{ding2024horizon} departed from these approaches, showing improvements by adopting a planning and lookahead based approach to language generation.

To the best of our knowledge, the existing body of work in FIM code generation has primarily focused on either pre-training base LLMs or exploring alternative architectures and training methodologies. A significant gap in the existing literature is the lack of focused investigation into the intrinsic FIM capabilities of instruction-tuned LLMs – models already adapted for following instructions.  Our work aims to bridge this gap by specifically evaluating and enhancing the FIM performance of models that have already been fine-tuned for instruction following, offering a novel perspective on leveraging these readily available and powerful models for this crucial code completion task.

\section{Conclusion}

Supervised fine-tuning considerably enhances the generation of code that can be evaluated directly, significantly diminishing the reliance on intricate post-processing. Our fine-tuned \texttt{Qwen2.5-Coder} models achieve substantial performance gains on the HumanEval Infilling and SAFIM benchmarks. This underscores targeted fine-tuning as a route to directly utilize raw LLM outputs.

\clearpage
\newpage
\section{Limitations}
First, our evaluation is primarily focused on the Python programming language, as reflected in the HumanEval Infilling and SAFIM benchmarks. The generalizability of our findings to other programming languages, which may exhibit different syntactic structures and code completion patterns, remains an open question. Future work should explore the application of our fine-tuning approach and the resulting reduction in post-processing needs across a more diverse set of languages.

Second, the instruction fine-tuning data we created, while effective, was generated using a specific LLM (\texttt{Mixtral-8x22B}) and a defined set of splitting strategies. The quality and diversity of this synthetic data directly influence the performance of our fine-tuned models. Exploring alternative data generation methods, incorporating human-annotated FIM examples, or scaling the size and diversity of the training data could potentially lead to further improvements in FIM generation and a more robust elimination of post-processing requirements.

Finally, our evaluation focused on specific benchmarks designed for FIM code generation. While these benchmarks are widely used, they represent a specific type of FIM task. The performance of our fine-tuned models and the necessity of post-processing might vary in more complex or less constrained FIM scenarios encountered in real-world code editing environments. Further investigation into the applicability of our findings to such diverse scenarios is warranted.




\bibliography{anthology,custom}
\bibliographystyle{acl_natbib}

\clearpage
\appendix


\begin{figure*}[ht!]
\centering

\begin{tcolorbox}[title={Supervised Finetuning Prompt for Fill-in-the-Middle Code Generation}, colback=red!0, left=2pt,right=2pt,top=0pt,bottom=0pt]

{ 
Split the provided Python code into three parts: (1) prefix, (2) middle, and (3) suffix. The split can be made at any character position. The "middle" section should be from one of the following categories.
\begin{enumerate}[leftmargin=*, itemsep=0pt]
    \item A random span
    \item An algorithmic block
    \item A control-flow expression
    \item An API function call
    \item An assignment expression
\end{enumerate}
Note that, when combined, the prefix, middle, and suffix must recreate the original code in its entirety.

\vspace{0.2cm}
The input code is as follows. \\
\verb|```|python \\
\{content\} \\
\verb|```|

\vspace{0.2cm}
Provide 5 examples of prefix, middle, and suffix in the following format. Additionally, label the middle span as one of the five categories listed above.

\vspace{0.2cm}
\# Example: example\_number

\vspace{0.2cm}
\#\# Prefix \\
\verb|```|python \\
\# your code here \\
\verb|```|

\vspace{0.2cm}
\#\# Suffix \\
\verb|```|python \\
\# your code here \\
\verb|```|

\vspace{0.2cm}
\#\# Middle \\
\verb|```|python \\
\# your code here \\
\verb|```|

\vspace{0.2cm}
\#\# Label \\
\vspace{0.1cm}
}
\end{tcolorbox}
\caption{Prompt template to generate fill-in-the-middle training samples.}
\label{fig:datagen}
\end{figure*}

\begin{figure*}[ht!]
\centering

\begin{tcolorbox}[title={Supervised Finetuning Prompt for Fill-in-the-Middle Code Generation}, colback=red!0, left=2pt,right=2pt,top=0pt,bottom=0pt]

{ 
You are given an incomplete code with a prefix and suffix. Your task is to generate the middle section.

\vspace{0.3cm}
\# Prefix \\
\verb|```|python \\
\{prefix\} \\
\verb|```| \\

\vspace{0.3cm}
\# Suffix \\
\verb|```|python \\
\{suffix\} \\
\verb|```| \\

\vspace{0.3cm}
\# Middle \\
\verb|```|python \\
\# your code here \\
\verb|```| \\

\vspace{0.3cm}
Middle section generation guidelines:
\begin{enumerate}[leftmargin=*, itemsep=0pt]
    \item The middle section must, when combined with the prefix and suffix, form a complete code without syntax errors. Ensure that the end of the middle section does not overlap with the start of the suffix.
    \item Do not include any explanations or notes.
\end{enumerate}
\vspace{0.1cm}
}
\end{tcolorbox}
\caption{Prompt template for supervised finetuning for fill-in-the-middle code generation.}
\label{fig:utg_prompt}
\end{figure*}

\begin{figure*}[t]
\centering
\begin{adjustbox}{valign=t,minipage=0.95\textwidth}
\begin{tabular}{l}
\lstset{escapechar=@,style=CustomPy,basicstyle=\fontsize{10}{15}\selectfont}
\begin{lstlisting}
def single_line_infill_postprocess(completion):
    lines = completion.splitlines()
    for line in lines:
        current_line = line.strip()
        if not current_line:
            continue
        if current_line.startswith("#"):
            continue
        return line
    return ""

def multi_line_infill_postprocess(completion, num_lines):
    assert num_lines > 0
    l = 0
    completion_lines = []
    for line in completion.split("\n"):
        completion_lines.append(line)
        current_line = line.strip()
        if current_line and not current_line.startswith("#"):
            l += 1
            if l == num_lines:
                break
    completion = "\n".join(completion_lines)
    return completion


def remove_overlap_prefix_middle(prefix, middle):
    prefix_len = len(prefix)
    middle_len = len(middle)
    for i in range(min(prefix_len, middle_len), 0, -1):
        if middle.startswith(prefix[-i:]):
            return middle[i:]
    return middle


def remove_overlap_middle_suffix(middle, suffix):
    suffix_len = len(suffix)
    middle_len = len(middle)
    for i in range(min(middle_len, suffix_len), 0, -1):
        if middle.endswith(suffix[:i]):
            return middle[:-i]
    return middle

def random_span_infill_postprocess(completion, prefix, suffix):
    completion = remove_overlap_prefix_middle(prefix, completion)
    completion = remove_overlap_middle_suffix(completion, suffix)
    return completion

\end{lstlisting}
\end{tabular}
\end{adjustbox}
\caption{
Post-processing functions for different HumanEval infilling tasks.
}
\label{fig:code_example}
\end{figure*}

\end{document}